\documentclass[journal]{IEEEtran}
\usepackage[numbers,sort&compress]{natbib}
\usepackage{underscore}
\usepackage{color}
\usepackage{float}

\usepackage{amsfonts}

\ifCLASSINFOpdf
   \usepackage[pdftex]{graphicx}
\else
\fi

\usepackage{algorithm}
\usepackage{algpseudocode}

\usepackage{url}
\begin{document}
\bibliographystyle{IEEEtr}

\title{\begin{huge}TDMA in Adaptive Resonant Beam Charging for IoT Devices\end{huge}}

\author{Mingliang~Xiong, Mingqing~Liu, Qingqing~Zhang,
        Qingwen~Liu*, Jun~Wu, and~Pengfei~Xia 
\thanks{* Corresponding
	author: Qingwen Liu.}

\thanks{
	M. Xiong, M. Liu, Q. Zhang, Q. Liu, J. Wu, P. Xia 
	are with the Department of Computer Science and Technology, Tongji University, Shanghai, People’s Republic of China, 
	(email: xiongml@tongji.edu.cn, 
	18392105294@163.com, 
	anne@tongji.edu.cn, 
	qliu@tongji.edu.cn, 
	wujun@tongji.edu.cn, 
	pengfei.xia@gmail.com).}}

\maketitle

\begin{abstract}
Resonant beam charging (RBC) can realize wireless power transfer (WPT) from a transmitter to multiple Internet of things (IoT) devices via resonant beams. The adaptive RBC (ARBC) can effectively improve its energy utilization. In order to support multi-user WPT in the ARBC system, we propose the time-division multiple access (TDMA) method and design the TDMA-based WPT scheduling algorithm. Our TDMA WPT method has the features of concurrently charging, continuous charging current, individual user power control, constant driving power and flexible driving power control. The simulation shows that the TDMA scheduling algorithm has high efficiency, as the total charging time is roughly half (46.9\% when charging 50 receivers) of that of the alternative scheduling algorithm. Furthermore, the TDMA for WPT inspires the ideas of enhancing the ARBC system, such as flow control and quality of service (QoS).

\end{abstract}

\begin{IEEEkeywords}
Resonant Beam Charging, Wireless Power Transfer, Time-Division Multiple Access, Internet of Things.
\end{IEEEkeywords}

%
\IEEEpeerreviewmaketitle

\section{Introduction}

	\IEEEPARstart{W}{ireless} power transfer (WPT) becomes an important and urgent need for mobile devices, such as smartphones~\cite{Carroll2010} and Internet of things (IoT) devices~\cite{Georgiou2017,Yu2016A}. Recently, Qihui Wu and Guoru Ding have made the primary contributions to cognitive Internet of things (CIoT) and its applications, which provide valuable references for this work~\cite{Ding2018An,Wu2014Cognitive}. For the emerging area of CIoT, a high-power and human-safe WPT can promote the massive data acquisition and analysis of CIoT deceives. WPT, as known as wireless charging, can be classified into non-radiative coupling-based charging and radiative radio frequency(RF)-based charging~\cite{Xiao2016}. In terms of power transmission distance, wireless charging can be classified into far-field charging and near-field charging~\cite{Garnica2013}. The WPT technologies, such as inductive coupling~\cite{Valtchev2012}, magnetic resonance coupling~\cite{Assawaworrarit2017,Cannon2009}, capacitive coupling~\cite{Kang2016}, radio waves~\cite{Shinohara2014}, and laser~\cite{Fakidis2018}, have been well investigated in researches. However, these WPT technologies are facing various technical challenges. For instance, inductive coupling has to operate in short-distance; radio-wave charging suffers low efficiency; and laser charging has the difficulty to meet the safety requirements.
	
\begin{figure}
	\centering
	\includegraphics[width=3.5in]{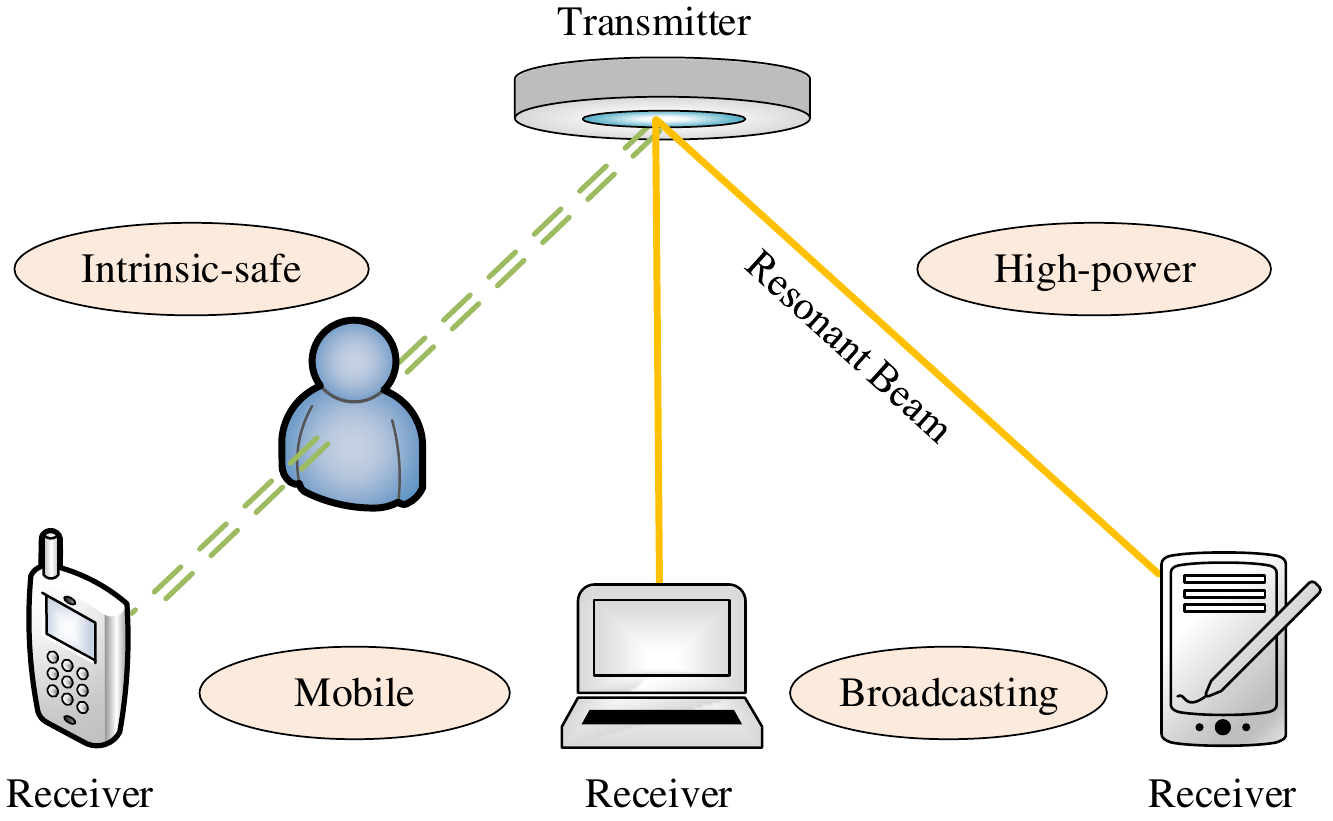}		
	\caption{Resonant Beam Charging Features}	
	\label{fig:DLC-sys}		
\end{figure}
	
	Resonant beam charging (RBC), as known as  distributed laser charging (DLC)~\cite{Qingwen2016}, can transfer energy via a resonant beam between the transmitter and the receiver. Fig.~\ref{fig:DLC-sys} shows the RBC features including high-power, intrinsic-safe, mobile, and broadcasting. RBC can safely charge multiple moving devices with Watt-level power over meter-level distances like WiFi communications. The adaptive resonant beam charging (ARBC) system proposed in \cite{Qingqing2017} improved the WPT efficiency based on feedback control like link adaptation in wireless communications.

	Although single-user ARBC was presented in \cite{Qingqing2017}, multi-user WPT in the ARBC system is important and worth of investigating for the applications like wireless sensor networks~\cite{MADHJA201589,Xu2016Efficient,Feng2017Locator}. Thus, multiple access in the ARBC system needs to be studied. Furthermore, it is also interesting to find out an efficient scheduling method for multi-user WPT in the ARBC system, in order to satisfy each user's energy requirement.

	In this paper, we propose two types of power control methods for the multi-user ARBC system: the alternative charging and the time-division multiple access (TDMA) charging. 
	
	\emph{Alternative charging}: The transmitter transfers proper power to one receiver for a period of time and then switches to another. However, this design has two obvious limitations: 1) the efficiency is low as only one receiver can be charged in each charging period; and 2) charging current may be discontinuous due to user switch.
	
	\emph{TDMA charging}: It combines time-division multiplexing (TDM) and pulse width modulation (PWM) for multi-user WPT, which has neither of the above limitations. In communication systems, TDM was originally introduced to combine multiple signals in one channel in time domain\cite{Gallagher2003A}. Then, TDM was adopted for WPT as TDM-WPT in~\cite{Jiang2017}. PWM technology provides an approach to convert pulses into continuous current, which is widely adopted in switching mode power supply~(SMPS)~\cite{Forsyth1998}. We employ TDM to handle multi-user WPT and employ PWM to realize continuous current charging.
	
	The contribution of this paper includes: 1)  We design and analyze the PWM-assisted TDMA method for multi-user WPT in the ARBC system; 2) We propose the TDMA-based scheduling algorithm to realize efficient and flexible multi-user WPT in the ARBC system. The proposed TDMA-based charging design has the following features:
	\begin{itemize}
		\item \emph{Concurrently charging}: Multiple receivers can be charged concurrently;
		\item \emph{Individual user power control}: The charging power of each receiver can be controlled independently;
		\item \emph{Continuous charging current}: The charging current  of each receiver is continuous;
		\item \emph{Constant driving power}: The driving power at the transmitter can keep constant and avoid being frequently switched as in alternative charging, which leads to low-complexity and robustness for the transmitter design;
		\item \emph{Flexible driving power control}: The driving power can be flexible regardless of charging power requirements.
	\end{itemize}
	Compared with the alternative charging method, the TDMA-based method has the following benefits:
	\begin{itemize}	
		\item \emph{Charging efficiency}: The total charging time for multiple users is 46.9\% of that of the alternative charging;
		\item \emph{Output capability}: There is no limit on output power;
    	\item \emph{User capacity}: The system can support more users due to the improvement of output capability.
	\end{itemize}

	In Section~\ref{sec:Alt-Cha-Met}, we introduce the principle of RBC; present the point-to-point ARBC system; depict the multi-user charging problem in the ARBC system; and demonstrate the alternative charging method. In Section~\ref{sec:TDMA}, we propose the TDMA method for multi-user WPT in the ARBC system. In Section~\ref{sec:algorithm}, we depict the TDMA-based scheduling algorithm. In Section~\ref{sec:sim-est}, we illustrate the simulation to compare the efficiency between the alternative and TDMA-based charging methods, and summarize the features of the TDMA-based scheduling algorithm. Finally, in Section~\ref{sec:conclusion}, we make conclusions and discuss the future research topics. 
	
\section{System Description}
\label{sec:Alt-Cha-Met}

	In this section, we firstly introduce the principle of resonant beam charging (RBC). Then, we depict the point-to-point adaptive resonant beam charging (ARBC) system which supports the time-division multiple access (TDMA) charging in Section~\ref{sec:TDMA}. Next, we demonstrate the necessity of multi-user power control when charging multiple receivers. At last, we  propose the alternative scheduling algorithm and discuss its features. 
	 
\subsection{Resonant Beam Charging}

\begin{figure}
	\centering
	\includegraphics[width=3.4in]{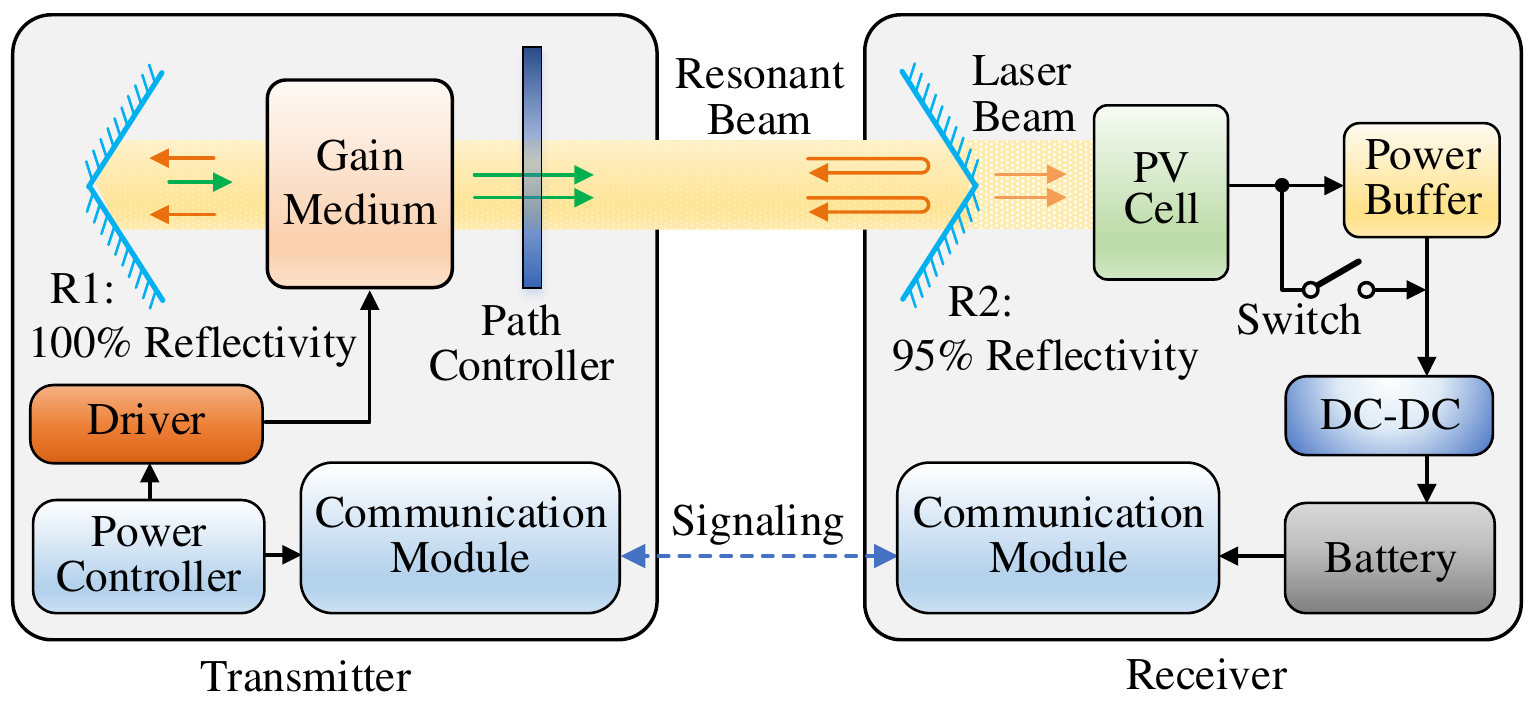}
	\caption{Adaptive Resonant Beam Charging  System}
	\label{fig:system-model}
\end{figure}

	As shown in Fig.~\ref{fig:system-model}, the RBC system consists of two parts: the transmitter and the receiver. In the system, the transmitter and the receiver constitute the resonant cavity. Therefore, differing from the traditional laser cavity, the RBC cavity is separated in space. R1 and R2 are retro-reflectors which can reflect the beam to the reverse direction regardless of the incident angle~\cite{Newman2012}. Between the two retro-reflectors, the photons are reflected repeatedly to stimulate out more photons while passing through the gain medium. Hence, the resonant beam connecting R1 and R2 can be formed naturally without concerning about the location of the receiver. The above mechanism guarantees the features of mobility and broadcasting. As a retro-reflector with 95\% reflective, R2 allows a part of the photons passing through. Therefore, a laser beam can output from the back of R2, which can provide high-power energy. Furthermore, any obstacles such as human bodies and cloths that enter the resonating beam path will disrupt the resonance immediately, which provides the intrinsic safety. At last, the beam can be converted into current by the photovoltaic (PV) cell to charge the battery.
	
\subsection{Point-to-point System}

	The point-to-point system is depicted in Fig.~\ref{fig:system-model}. In the transmitter, the driver provides the driving power for the gain medium to generate the beam. This process converts electrical energy into optical energy. Then, the beam travels through the air, resonating between the transmitter and the receiver. The path controller is composed of an optical switches array, and it is adopted to switch the on-off state of the beam path. In the receiver, the PV cell converts optical energy into electrical energy. As the most important element for the implementation of the TDMA charging method, the power buffer can convert PWM power into continuous smooth power. However, for the alternative charging, the power buffer should be bypassed by closing the switch. The direct current to direct current (DC-DC) converter ensures that the charging voltage and the charging current can satisfy the battery's requirement. The communication modules can transmit battery states from the receiver to the transmitter via the signaling channel. At the transmitter, the power controller can control the driving power.

	In the system model, the electrical energy provided by the driver is converted into optical energy with the	 efficiency $\eta_{\rm s}$. The transmission efficiency of the path controller is $\eta_{\rm pc}$.  Due to propagation attenuation, the resonant beam power experiences power loss with the transmission efficiency $\eta_{\rm t}$. Moreover, the photoelectric conversion efficiency of the PV cell is $\eta_{\rm r}$. The conversion efficiency of the power buffer is $\eta_{\rm pb}$. The conversion efficiency of the DC-DC converter is $\eta_{\rm dc}$. We define the driving power as $P_{\rm d}$, then the charging power $P_{\rm c}$ can be obtained as:
	 \begin{equation}
	 P_{\rm c}=\eta_{\rm s}\eta_{\rm pc} \eta_{\rm t} \eta_{\rm r} \eta_{\rm pb} \eta_{\rm dc} P_{\rm d}.
	 \label{equ:Pc=eta*Ps}
	 \end{equation}

\begin{figure}
	\centering\includegraphics[width=2.8in]{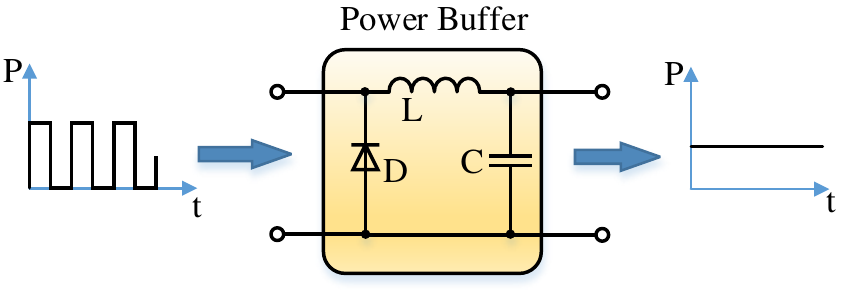}
	\caption{Power Buffer Principle}
	\label{fig:power-buffer}
\end{figure}

	The principle of the power buffer is depicted in Fig.~\ref{fig:power-buffer}. The PWM wave is input to the power buffer, and the output is the continuous current. The circuit of the power buffer is similar to that of a switching mode power supply (SMPS), which consists of the energy storage components such as inductors and capacitors~\cite{Weaver2006}. Inductors have the ability to store electrical energy by converting it into magnetic energy. Similarly, capacitors store electrical energy in the form of electric charges. As pulses enter the power buffer, the analysis can be divided into two stages: 
	\begin{itemize}
		\item ON stage: The input power of the power buffer is at a high level, which is notated as the  peak input power $P_{\rm bip}$; and the diode remains a cut-off state while the capacitor is being charged by the inductor. As a result, the output voltage rises up slowly.
		\item OFF stage: Following the ON stage, the input power turns to zero, and the diode is switched on. Thus, the energy stored in the inductor is released to the capacitor and the load. Therefore, the output voltage goes down slowly.
	\end{itemize}

	In theory, there is no energy consumption in such components as inductors and capacitors without resistance.  Because of the law of energy conservation in a period:
	\begin{equation}
	E_{\rm bo}=E_{\rm bi},
	\end{equation}
	it can be induced that:
	\begin{equation}
	P_{\rm bo} T_{\rm w} = P_{\rm bip} \cdot T_{\rm ON} + 0\times(T_{\rm w}- T_{\rm ON}),
	\label{equ:Ei=Eo}
	\end{equation}
	where $E_{\rm bi}$ is the input energy, $E_{\rm bo}$ is the output energy, $P_{\rm bip}$ is the peak power of the input pulse, $P_{\rm bo}$ is the constant output power, $T_{\rm ON}$ is the time width of the pulse (ON stage), and $T_{\rm w}$ is the period of the PWM wave. The duty cycle $\delta$ of the PWM wave is depicted as:
	\begin{equation}
	\delta = \frac{T_{\rm ON}}{T_{\rm w}}.
	\label{equ:delta}
	\end{equation}
	Hence, from (\ref{equ:Ei=Eo}) and (\ref{equ:delta}), it can be derived that:
	\begin{equation}
	P_{\rm bo}=\delta P_{\rm bip} .
	\label{equ:Po=deltaPp}
	\end{equation}
	Then, from~(\ref{equ:Pc=eta*Ps}) and (\ref{equ:Po=deltaPp}), the charging power can be obtained as:
	\begin{equation}
	P_{\rm c}=\eta \delta P_{\rm d},
	\label{equ:tdmpwm-Psp-to-Pc}
	\end{equation}
	where
	\begin{equation}
	\eta=\eta_{\rm s}\eta_{\rm pc} \eta_{\rm t} \eta_{\rm r} \eta_{\rm pb} \eta_{\rm dc} .
	\label{equ:efficiency}
	\end{equation}

	From (\ref{equ:tdmpwm-Psp-to-Pc}), we can verify that the charging power can be determined by the duty cycle of the PWM beam.
\subsection{Multi-user Power Control}

\begin{figure}
	\centering
	\includegraphics[width=3.5in]{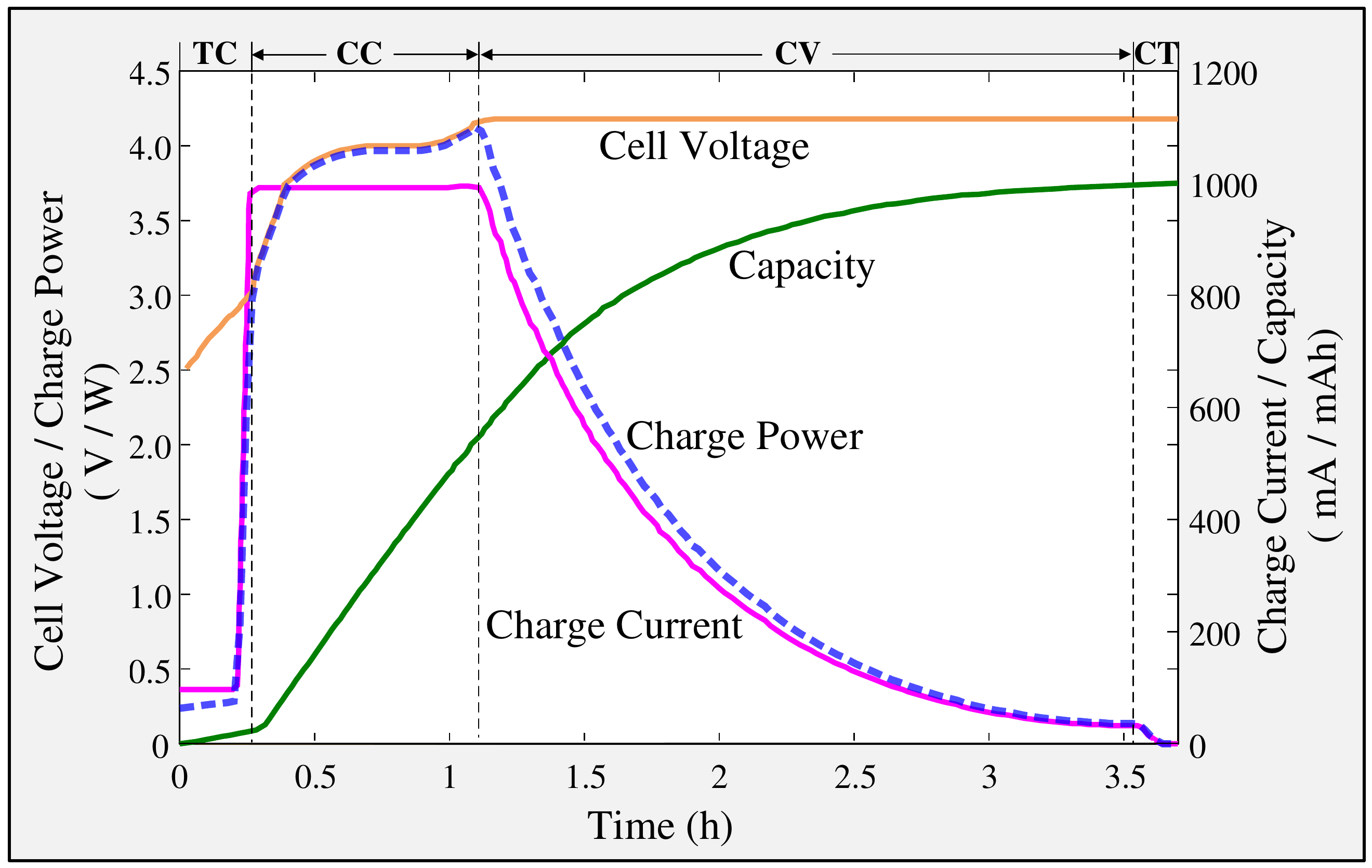}
	\caption{Li-ion Battery Charging Profile}
	\label{fig:li-battery-profile}
\end{figure}
\begin{figure}
	\centering
	\includegraphics[width=3.3in]{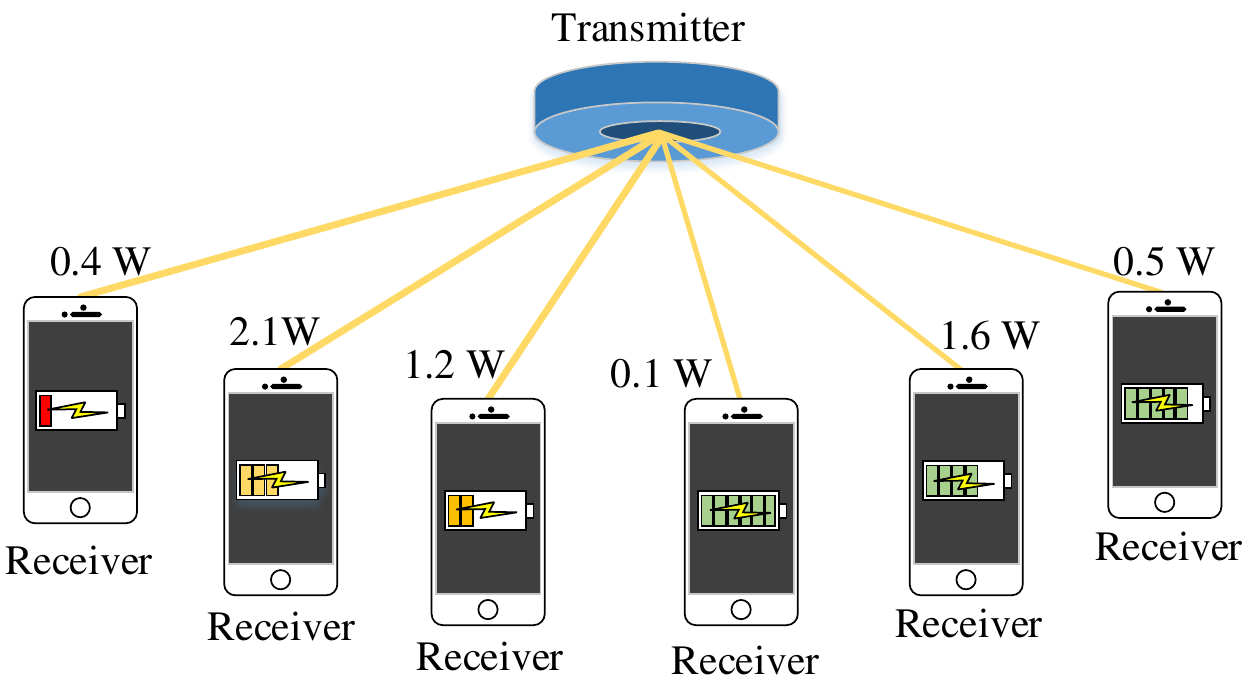}
	\caption{Multi-user Power Control}
	\label{fig:multi-user}
\end{figure}

	RBC has the broadcasting feature, as the resonant beams can be formed between one transmitter and multiple receivers. However, the broadcast charging is less efficient in energy utilization, because the RBC system can not track the receivers' power requirements concurrently.
	
	Owing to the advantages of high energy density and long cycling life, Li-ion batteries are the most common energy storage components  in mobile devices. The specified constant current constant voltage (CC-CV) charging algorithm can optimize the battery performance~\cite{Dearborn2005}. The CC-CV algorithm can be depicted in Fig.~\ref{fig:li-battery-profile} which is called the Li-ion battery charging profile. The charging profile has four stages in a charge circle, which are trickle charge (TC), constant current (CC), constant voltage (CV), and charge terminal (CT) respectively~\cite{Dearborn2005}. 
	
	According to the CC-CV algorithm, the battery is charged with the dynamic power in the ARBC system. From Fig.~\ref{fig:li-battery-profile}, when charging the 1000 mAh Li-ion battery, the charging power rises up from zero to the maximum value~(4.2W) and then goes back to zero smoothly. Hence, each receiver has a unique desired charging power $P_{\rm c}$ according to the charging profile during the charging period. As shown in Fig.~\ref{fig:multi-user}, the charging power delivered to each receiver should be specified. In order to optimize each receiver's battery performance, the charging power per receiver should be controlled according to its battery charging profile. Therefore, how to realize multi-user power control in the ARBC system to track the receivers' requirements concurrently becomes an important issue.
	
\subsection{Alternative Charging}
\begin{figure}
	\centering
	\includegraphics[width=3.3in]{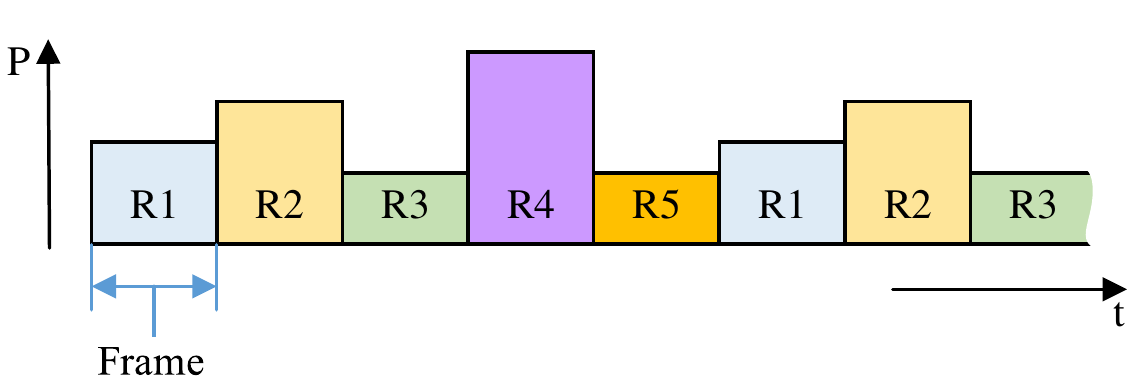}
	\caption{Alternative Charging Principle}
	\label{fig:alt-cha-met}
\end{figure}

	In order to realize multi-user power control, the alternative charging is an intuitive method without the power buffer (i.e., the switch is closed in Fig.~\ref{fig:system-model}). In Fig.~\ref{fig:alt-cha-met}, time is divided into a sequence of small frames with the same width. In each frame, the transmitter generates a beam for only one receiver (R1, R2, R3, R4 or R5) sequentially with the receiver's battery desired charging power. This charging procedure (charging from R1 to R5 in turns) will repeat until all the receivers are fully charged. Thus, the scheduling algorithm can be depicted as:
	
    \begin{itemize}  
	\item  \emph{Alternative scheduling algorithm}: To charge the receivers sequentially with their desired battery charging power, repeatedly, and removes those receivers whose capacity reaches to 100\%.
	\end{itemize}
	
	However, after being charged in one frame, the receiver have to wait for its next charging frame in a long time (e.g., four frames in Fig.~\ref{fig:alt-cha-met}). Therefore, the alternative charging has the disadvantages of low efficiency and  discontinuous battery charging current.
	
	The driving power at the transmitter is switched over a frame. The frequent driving power adjustment may be harmful for the gain medium and lead to the complex transmitter design. Moreover, the driving power is constrained to satisfy only one receiver's requirement per frame, which is inflexible and leads to the restricted output capability. To solve those problems, a more efficient and flexible multiple access method will be proposed in the next section.

\section{TDMA Charging Design}
\label{sec:TDMA}

	In this section, we propose the TDMA charging design in the ARBC system, which have the features of currently multi-user charging, individual user power control and continuous charging current. We at first demonstrate the TDMA charging principle. Then, we propose the digitalized design to facilitate the computer control.
	
\subsection{TDMA Charging Principle}

\begin{figure}
	\centering
	\includegraphics[width=3.5in]{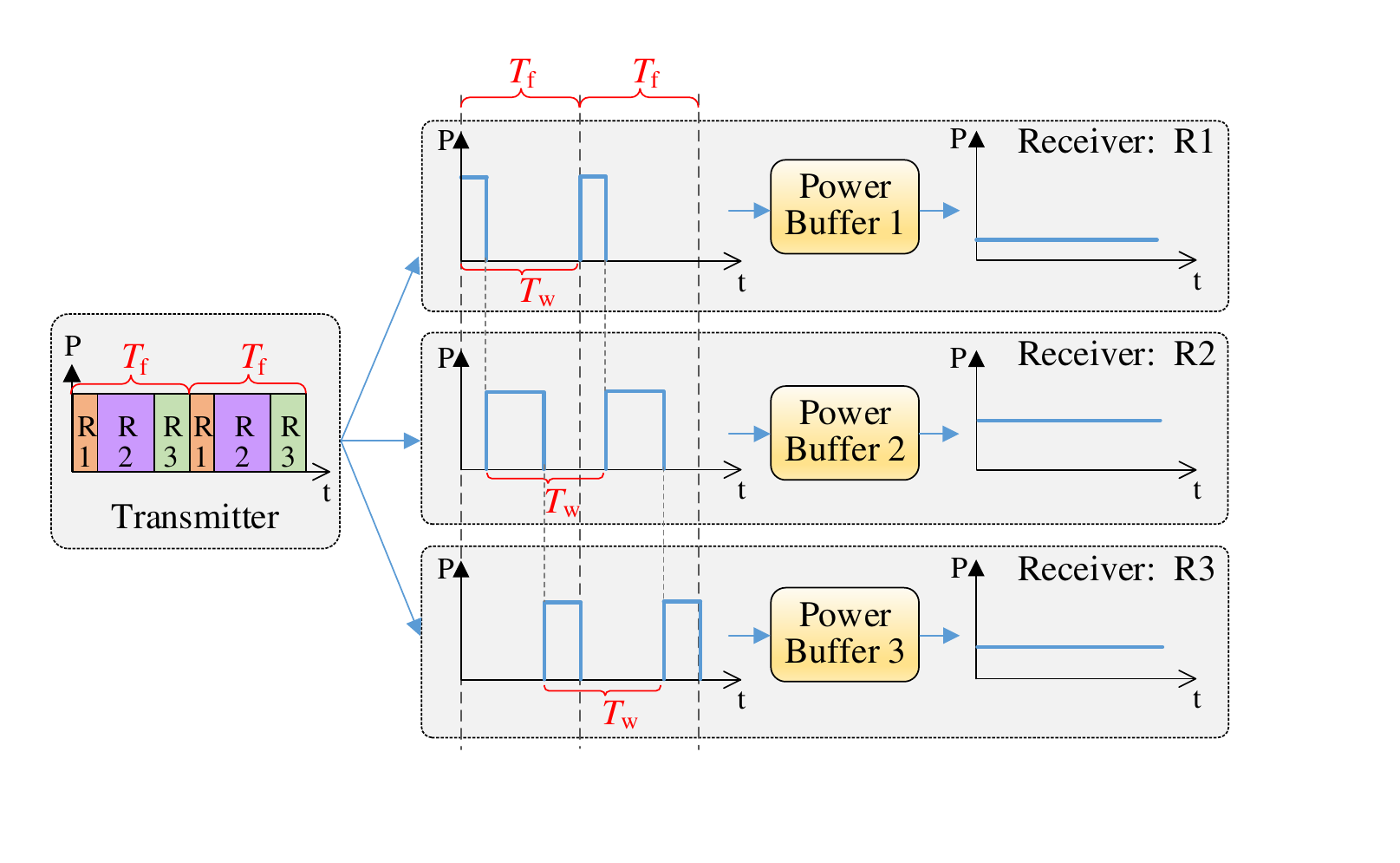}
	\caption{Time-Division Multiple Accessing Charging Principle}
	\label{fig:tdm-pwm-met}
\end{figure}

	With the system illustrated in Section~\ref{sec:Alt-Cha-Met}, the transmitter should send the ON-OFF alternative PWM beam to the receiver, and the charging power can be determined by the duty circle (pulse width) of the PWM beam. The pulse takes only two stages: the ON stage and the OFF stage. During the ON stage, the transmitter opens the path for the beam orientating to one receiver, and closes during the corresponding OFF stage. In the OFF stage the transmitter is idle without energy output. If the second receiver's ON stage duration is shorter than the first receiver's idle time (the OFF duration), the transmitter could send the beam pulse to the second receiver in the first receiver's idle time, in order to improve the time utilization.

	Based on the above idea, we can borrow the TDMA principle to combine multiple beam pulses in one frame. In Fig.~\ref{fig:tdm-pwm-met}, in a frame period $T_{\rm f}$, the transmitter generates three beam pulses one-by-one to three receivers (R1, R2 and R3) with different pulse width for each receiver. Then, in the following frames, the transmitter repeats this beam pulses pattern. As a result, from each receiver's perspective, a PWM wave with the period of $T_{\rm w}$ is received and eventually converted by the power buffer to a continuous  power. However, from the transmitter's point of view, it sends three PWM beams to three receivers. These PWM beams take the same fixed transmitting power and the same period $T_{\rm w}$ with different pulse width per receiver. Because the pulse widths of the three PWM beams are different, the output powers from the power buffers are distinct. Actually, the width of a frame is equal to the period of each PWM beam, namely
	\begin{equation}
		T_{\rm w}= T_{\rm f}.
		\label{equ:Tw=Tp}
	\end{equation}
	
	From~(\ref{equ:tdmpwm-Psp-to-Pc}), the transmitter can calculate the desired duty circle $\delta$ for each receiver according to its desired charging power $P_{\rm c}$ (as shown in Fig.~\ref{fig:li-battery-profile}). If $\delta$ and $T_{\rm w}$ are determined, the desired pulse width $T_{\rm ON}$ can be obtained from (4). Then, $N_{\rm c}$ receivers can be selected to form a frame, according to the \emph{selecting rule} that the sum of these receivers' desired pulse widths must be less than a frame time:
	\begin{equation}
	\sum\limits_{i=0}^{N_{\rm c}}{T_{\rm ON_\mathit{i}}} \le T_{\rm f},
	\label{equ:select-law}
	\end{equation}
	where $T_{\rm ON_\mathit{i}}$ is the time width of the pulse sent to the $i\mbox{-th}$ selected receiver.
	
	The transmitter can send beam pulses to each selected receiver with corresponding pulse width and order, and all of the selected receivers can obtain continuous smooth current concurrently. 

	In summary, the TDMA charging originates from the concepts combination of time-division multiplexing (TDM) and PWM. The advantages includes the following three aspects:
	\begin{enumerate}
		\item Multiple receivers can be charged with different charging power continuously and concurrently;
		\item The transmitter generates beams to receivers with the same driving power;
		\item The driving power can be customized to an desired value flexibly.
	\end{enumerate}

\subsection{Digitalized TDMA Charging Design}

\begin{figure}
	\centering
	\includegraphics[width=3.5in]{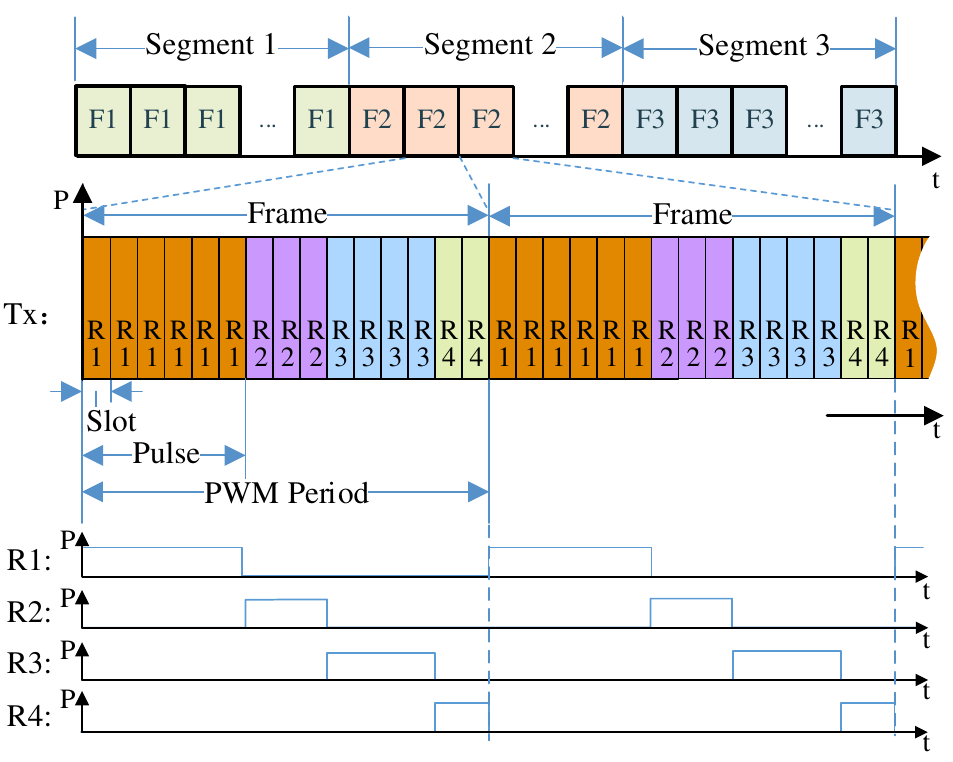}
	\caption{Digitalized TDMA Charging Design}
	\label{fig:dig-tdm-wpm}
\end{figure}

	The TDMA charging is to combine the PWM pulses for the selected receivers in a frame. In order to adopt the computer control in the TDMA charging, we design the digitalized rule which contains three structures: segment, frame, and slot. 

	Fig.~\ref{fig:dig-tdm-wpm} shows the digitalized TDMA charging design. The time is divided into a number of segments. Each segment contains several identical frames. Each frame is divided into a number of slots. The slots are the identical time units and can be allocated to some proper receivers. For example, in Fig.~\ref{fig:dig-tdm-wpm}, the notations R1, R2, R3, and R4 represent the receivers. If one slot is allocated to a receiver, then the transmitter can generate the beam to this receiver in this time slot. In Fig.~\ref{fig:dig-tdm-wpm}, the adjacent slots with the same receiver notation constitute this receiver's  pulse. The pulses marked with the same receiver notation constitute a PWM that will be sent to the corresponding receiver. The peak power of these pulses is constant. For instance, in Fig.~\ref{fig:dig-tdm-wpm}, the first 6 slots in each frame F2 are marked as R1, so the transmitter Tx will switch the beam orientating to the receiver R1 during those time slots, as a result, R1 will receive a PWM wave whose pulse width and period is equal to~6 slots and~15 slots respectively. 

	All of the slots have the same time width $T_{\rm s}$. The time width of each frame is $T_{\rm f}$. Hence, the total slot number, $N_{\rm s}$, in one frame can be obtained as: 
	\begin{equation}
	N_{\rm s} = \frac{T_{\rm f}}{T_{\rm s}} .
	\end{equation}
	For instance, Fig.~\ref{fig:dig-tdm-wpm} shows the details of the frame F2 which is divided into $N_{\rm s}=15$ slots. 
	
	As the peak power, $P_{\rm p}$, of the  pulse beam is constant, the driving power~$P_{\rm d}$ in ON stage can also be a fixed value. However, the desired charging powers, $P_{\rm c}$, of the receivers are distinct.  From~(\ref{equ:tdmpwm-Psp-to-Pc}), given the value of $P_{\rm c}$, the desired PWM duty cycle $\delta$ can be induced as:
	\begin{equation}
	\delta=\frac{P_{\rm c}}{\eta P_{\rm d}}.
	\label{equ:delta2}
	\end{equation}
	In order to form a pulse for a specific receiver with a desired width, the transmitter needs to allocate the desired number of slots to the receiver. The desired slot number $N_{\rm c}$ can be induced as:
	\begin{equation}
	N_{\rm c}=\lceil \delta N_{\rm s} \rceil  = \lceil \frac{P_{\rm c}}{\eta P_{\rm d}} N_{\rm s} \rceil ,
	\label{equ:Nexp}
	\end{equation}
	where the ceiling notation is used to ensure that $N_{\rm c}$ is the proper integer which can provide enough energy for the receiver. For example, in Fig.~\ref{fig:dig-tdm-wpm}, $N_{\rm c}$ of the receivers from R1 to R4 are 6, 3, 4, and 2 respectively.

	Segment is a period of time $T_{\rm g}$, consisting of consecutive identical frames. However, the frames in different segments may not be identical. For example, in Fig.~\ref{fig:dig-tdm-wpm}, segment~1 consists of several frames F1, while in segment~2 the frames are F2. 
	
	The structure of segments is important in the design. The pulses for multiple receivers are arranged closely in a frame to ensure the optimal time distribution. If the pulse width of one receiver increases, the pulses of other receivers in the same frame must move their positions backward, which brings an unpredictable situation that the last pulse in the fame may have no enough place. To avoid the unpredictable situation, the pulse width of each PWM wave must be constant, and the position of each pulse in the frame must be fixed. Therefore, the segment structure should contain several identical frames to provide a long period of time for stably outputting PWM waves.
	
	In summary, the segment structure ensures the selected receivers to be charged continuously for an effective long duration. The segment is an important term in the \emph{OUTPUTTING} procedure of the TDMA scheduling algorithm, which will be specified in Section~\ref{sec:algorithm}.
	
	The above multiplexing of PWM waves leads to the concurrently charging feature. Hence, we define 
	\begin{itemize}
		\item \emph{Multiplexing numbe\textsf{}r}: $\psi$, as the number of the selected receivers in a frame.
	\end{itemize}
	 For instance, in Fig.~\ref{fig:dig-tdm-wpm}, $\psi$ of frame F2 is 4, because four receivers are contained in the frame. We define the average multiplexing number, $\Psi$, of a charging process as:
	\begin{equation}
	\Psi=\sum\limits_{0\le~t_i<T_{\rm u}}{\frac{\psi_i T_{\rm f}}{T_{\rm u}}}~~~(i=1,2,3,\ldots,n),
	\end{equation}
	where $\psi_{\rm i}$ is the multiplexing number of the $i$-th frame, $t_{i}$ is the time when starting to handle the $i$-th frame, and $T_{\rm u}$ is the time costed in the total charging process. The value of $\Psi$ can be used to evaluate the efficiency of the TDMA scheduling algorithm.
	
	In summary, the transmitter can control the charging power for each receiver by controlling the slot number allocated to the receiver. Therefore, the transmitter can switch the beam orientation toward the receiver according to its allocated slots in a frame. Based on the repeated frame structure per segment, the PWM wave can be obtained per receiver. Finally, relying on the power buffer, the receiver can be charged with continuous and smooth current.
	
\section{TDMA-based Scheduling Algorithm}
\label{sec:algorithm}

\begin{figure}
	\centering
	\includegraphics[width=3.5 in]{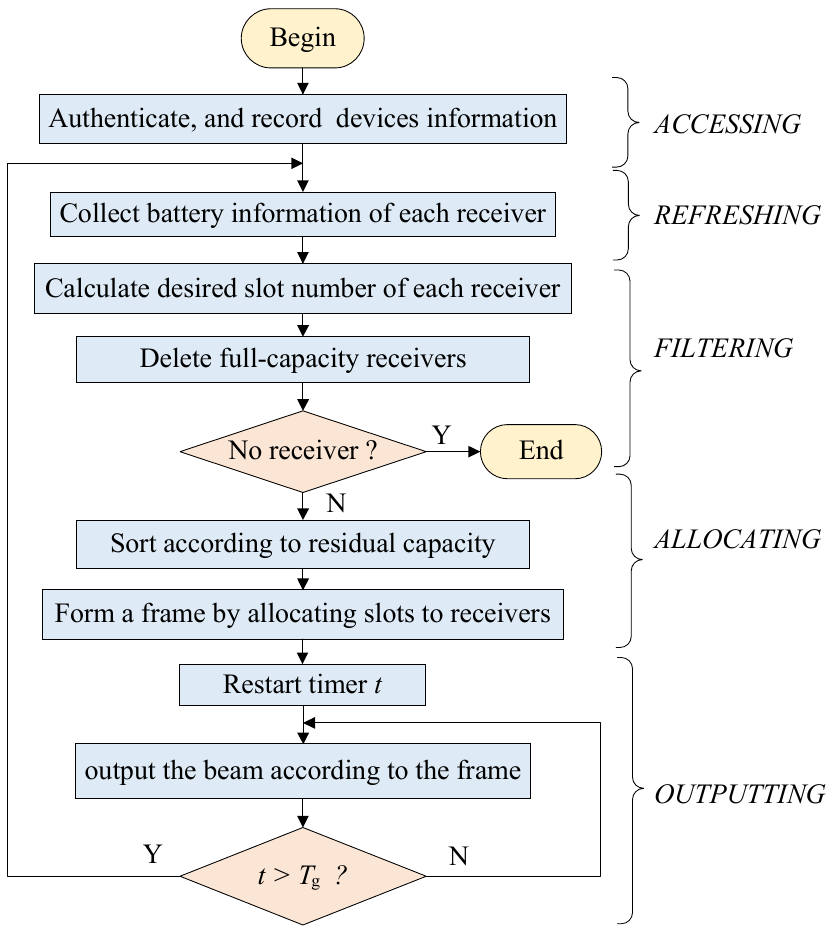}
	\caption{Flow Chart of TDMA Charging Algorithm}
	\label{fig:tdm-pwm-sche}
\end{figure}

    The above sections introduce the hardware requirements and the TDMA charging design. In this section, we will introduce the TDMA-based scheduling algorithm, which contains the rule of slots allocation and provides a logical scheduling procedure. In summary, we define
    \begin{itemize}  
    \item  \emph{TDMA scheduling algorithm}: To allocate slots to the receivers according to their desired charging power and residual battery capacity; then, form the frame structure, and operate a segment in a scheduling period.
	\end{itemize}

    As is depicted in Fig.~\ref{fig:tdm-pwm-sche}, we divide the TDMA scheduling algorithm into the following five parts:
    \begin{enumerate}
      \item     \emph{ACCESSING}: To send acknowledgements to the receivers after receiving the requests; then handle authentications and record the information of the accessed receivers;

      \item     \emph{REFRESHING}: To obtain the battery states of the receivers, such as residual battery capacity and desired charging power, via the signalling channels; and calculate the desired slots number of the receivers;

      \item     \emph{FILTERING}: To remove the information of the full-capacity receivers from the memory to ensure that those receivers are no longer charged;

      \item     \emph{ALLOCATING}: To sort receivers according to the residual battery capacity in ascending order; select some receivers with proper rules; and form a frame by allocating slots to the selected receivers;

      \item     \emph{OUTPUTTING}: To operate a segment of charging by running a loop, where the transmitter output beams according to the frames with a constant driving power.
    \end{enumerate}

\begin{table}[!t]
	\normalsize 
	\centering
	\caption{Device Profile Structure}
	\renewcommand{\arraystretch}{1.2}
	\centering
	\begin{tabular}{ l l }
		\hline
		Parameter & Description \\
		\hline
		$C_{\rm r}$ & residual battery capacity \\ 
		$P_{\rm c}$ & desired charging power\\
		$N_{\rm c}$ & desired slot number\\
		...& other parameters: ID, location, etc.\\
		\hline
	\end{tabular}
	\label{tbl:dev-prof}
\end{table}

\begin{table}[!t] 
	\normalsize
	\centering
	\caption{Data Notation Definitions}
	\renewcommand{\arraystretch}{1.2}
	\centering
	\begin{tabular}{ l l l}
		\hline
		Notation & Type & Description \\
		\hline
		$\mathbf{m}[~]$ & device profile array &  receivers to be accessed\\ 
		$\mathbf{\mathbf{S}}[~]$ & device profile array &  accessed receivers\\
		$\mathbf{F}[~]$ &  device profile array &   slot allocation \\
		$t$ & timer &  counting the time\\
		$N_{\rm s}$ & constant & slot number per frame\\
		$P_{\rm d}$& constant & driving power\\
		$T_{\rm s}$ & constant & slot width\\
		$T_{\rm g}$ & constant &   segment width \\
		$C_{\rm full}$ & constant & full battery capacity \\
		\hline
	\end{tabular}
	\label{tbl:data-def}
\end{table}

	The pseudocode of the TDMA scheduling algorithm is listed in Algorithm~\ref{alg:tdmpwm-charging}. We firstly introduce the data structure of the device profile, which records the information of the receiver. The parameters in the device profile structure are listed in Table~\ref{tbl:dev-prof}. The notation definitions of the data in the algorithm are listed in Table~\ref{tbl:data-def}. In the pseudocode, a pair of square brackets wrapping up a number ($i$, for example) after a symbol means an element (the $i$-th element, for example) of the array that marked by the symbol.

\begin{algorithm}[!t]
	
	\caption{TDMA Scheduling}\label{alg:tdmpwm-charging}
	\begin{algorithmic}[1]
		
		\State \textbf{begin}
		
		\For{$i$ = 1 to $N$}
		\label{alg:tdmpwm:acc-begin}
		\If {$\mathbf{m}[i]$  is authenticated } 
		\State $append ~ \mathbf{m}[i]  ~to ~\mathbf{S}$ 
		\EndIf
		\EndFor
		\label{alg:tdmpwm:acc-end}
		
		\For {$i$ = 1 to \emph{length}($\mathbf{S}$)}
		\label{alg:tdmpwm:ref-begin}
		\State \emph{refresh} $\mathbf{S}[i].C_{\rm r}$ and $\mathbf{S}[i].P_{\rm c}$
		\State calculate $\mathbf{S}[i].N_{\rm c}$ according to formula~(\ref{equ:Nexp})
		\State \textbf{If} { $\mathbf{S}[i].C_{\rm r} \geq  C_{\rm full}$ } \textbf{then}
		\emph{delete} $\mathbf{S}[i]$ \label{alg:tdmpwm:filtering}
		
		\EndFor		
		
		\If{$\mathbf{S}$ is empty}
		\State \textbf{end}
		\EndIf
		
		\label{alg:tdmpwm:ref-end}
		
		\State sort $\mathbf{S}$ according to $\mathbf{S}.~C_{\rm r}$ in ascending order			
		\label{alg:tdmpwm:sel-begin}
		
		
		\For{$i$ = 1 to \emph{length}($\mathbf{S}$)}
		
		\If{$\mathbf{S}[i].N_{\rm c} \leq N_{\rm s}-length(\mathbf{F})$}
		\For{$j$ = 1 to $\mathbf{S}[i].N_{\rm c}$}
		\State  \emph{append} $\mathbf{S}[i]$ \emph{to} $\mathbf{F}$
		\EndFor
		\EndIf
		
		\EndFor		\label{alg:tdmpwm:sel-end}
		
		\State restart timer $t$			\label{alg:tdmpwm:loo-begin}
		
		\While{$t \le T_{\rm g}$}
		\For{$i$ = 1 to \emph{length}($\mathbf{F}$)} 
		\State output to $\mathbf{F}[i]$ for $T_{\rm s}$ secs with driving power $P_{\rm d}$ 
		\EndFor
		\State sleep for $(N_{\rm s}-length(\mathbf{F}))\times T_{\rm s}$ secs without output
		\EndWhile		
		
		\State empty $\mathbf{F}$, and \textbf{goto} step~\ref{alg:tdmpwm:ref-begin}
		\label{alg:tdmpwm:loo-end}
	\end{algorithmic}
\end{algorithm}

	In the \emph{ACCESSING} procedure (step~\ref{alg:tdmpwm:acc-begin}-\ref{alg:tdmpwm:acc-end} in Algorithm~\ref{alg:tdmpwm-charging}), there are $N$ receivers waiting for connection. The transmitter handles the requests from the receivers $\mathbf{m}[i],(i=1,2,3,\ldots,N)$. If one receiver succeeds in authentication, a connection will be created and the device profile will be recorded as an element in the array $\mathbf{S}$. The instruction $append~\mathbf{m}[i]~to ~\mathbf{S}$ means adding an element which is a copy of $\mathbf{m}[i]$ to the tail of $\mathbf{S}$.

	In the \emph{REFRESHING} procedure (step~\ref{alg:tdmpwm:ref-begin}-\ref{alg:tdmpwm:ref-end} in Algorithm~\ref{alg:tdmpwm-charging}), the instruction \emph{refreshing} indicates that the transmitter sends notifications to the receivers in the array $\mathbf{S}$ and then collects the battery states returned via the signalling channels. One parameter of the battery state is the residual capacity $C_{\rm r}$ which implies the urgency level for charging. Another important parameter is the desired charging power $P_{\rm c}$ which determines the number of the slots that should be allocated. Then, it calculates the desired slot number $N_{\rm c}$ for each receiver based on~(\ref{equ:Nexp}). All the device profiles of the receivers will be refreshed in this procedure. The instruction $length(\mathbf{S})$ represents the number of the elements recorded in $\mathbf{S}$.
	
	In the \emph{FILTERING} procedure (step~\ref{alg:tdmpwm:filtering} in Algorithm~\ref{alg:tdmpwm-charging}), some of the receivers in the memory will be deleted. Only the devices which are not fully charged (the capacity is less than the maximum capacity $C_{\rm full} $) will remain in the array $\mathbf{S}$. The instruction \emph{delete} $\mathbf{S}[i]$ means removing the $i$-th element of the array $\mathbf{S}$. After the \emph{FILTERING} procedure, if $\mathbf{S}$ is found empty, the scheduling procedure will finish.

	The TDMA scheduling algorithm employs the digitalized TDMA charging rule. It divides a frame $T_{\rm f}$ into the fixed number $N_{\rm s}$ slots, with the fixed slot time-width $T_{\rm s}$. As in Section~\ref{sec:TDMA}, the frame width is equal to the period of the PWM wave. Hence, the frequency of each PWM wave is $1/T_{\rm f}$. \emph{ALLOCATING} is the procedure to form the frame by allocating the slots in a frame to the selected receivers.
	 
	In the \emph{ALLOCATING} procedure (step~\ref{alg:tdmpwm:sel-begin}-\ref{alg:tdmpwm:sel-end} in Algorithm~\ref{alg:tdmpwm-charging}), the array $\mathbf{S}$ is sorted in ascending order according to the parameters of the residual capacity $C_{\rm r}$, which means the receivers with low $C_{\rm r}$ will be arranged in the front of the array. Then, the allocating procedure begins. An allocating loop is employed to scan the array $\mathbf{S}$ from the head element to the tail. During the scanning, if $N_{\rm c}$ of the scanned receiver is less than the number of the unallocated slots in the frame (i.e. the condition in~(\ref{equ:select-law}) is satisfied), the receiver will be selected, and the corresponding slots in the frame will be allocated to this receiver. In a nutshell, the receiver with less residual capacity has the higher priority to be selected.
	
	In the scanning loop, the selected receivers are allocated with proper number of slots which occupy corresponding positions in the array $\mathbf{F}$ ( represents a frame), i.e., the elements in $\mathbf{F}$ represent the slots in the frame. If the receiver $\mathbf{S}[i]$ can be allocated with $N_{\rm c}$ slots, then $N_{\rm c}$ copies of the device profile recorded in $\mathbf{S}[i]$ will be appended, as the elements, to the tail of $\mathbf{F}$, except for the case that the frame cannot provide enough unallocated slots for the receiver.
	
	In summary, in the \emph{ALLOCATING} procedure, the group of the selected receivers is determined by both $C_{{\rm r}}$ and $N_{\rm c}$ of these receivers. However, the parameters are changed in each \emph{REFRESHING} procedure. Therefore, the number of the selected receivers is not fixed. As a result, the multiplexing number $\psi$ is dynamic during the whole charging process.
	
	At last, in the \emph{OUTPUTTING} procedure (step~\ref{alg:tdmpwm:loo-begin}-\ref{alg:tdmpwm:loo-end} in Algorithm~\ref{alg:tdmpwm-charging}), a timer $t$ is enabled firstly to count the elapsed time. Then, the procedure executes a charging loop to operate a segment, of which the total loop time is a segment time $T_{\rm g}$. Each period of the charging loop operates a frame, i.e., the transmitting beam is generated for the selected receivers from the array $\mathbf{F}$. Therefore, the timing of sending beams is  accordance with the elements in $\mathbf{F}$. The transmitter should send beam to the receiver recorded by the element. If the scanning finishes, the procedure will sleep for $(N_{\rm s}-length(\mathbf{F}))\times T_{\rm s}$~secs (the time of unallocated slots) without output. The sleep operation ensures that the period of the PWM waves equal to the frame width $T_{\rm f}$. The operation time of each output or sleep is one slot time $T_{s}$. Moreover, the driving power~$P_{\rm d}$ is also a constant value. If the charging loop finishes, all the elements in $\mathbf{F}$ will be deleted, and the procedure will jump to step~\ref{alg:tdmpwm:ref-begin} to start a new scheduling period.

\section{Simulation and Analysis}
\label{sec:sim-est}
	In this section, we introduce the MATLAB simulation of the alternative and TDMA scheduling algorithms. In the analysis, we focus on two aspects: 1) the comparison of the charging efficiency between the two algorithms; and 2) the features of the TDMA scheduling algorithm.  
	
\subsection{Simulation Parameters}
	In Section~\ref{sec:Alt-Cha-Met}, there are several conversion and transmission efficiencies in the ARBC system. The system components of the two scheduling algorithms are identical. We assume 
	\begin{itemize}
		\item the electro-optical conversion efficiency $\eta_{\rm s}$ is 40\%~\cite{810nmTransmitter,zhang2017distributed};
		\item the transmission efficiency over the air $\eta_{\rm t}$ is 100\%~\cite{Salman2009};
		\item the photoelectric conversion efficiency $\eta_{\rm r}$ is 50\% (GaAs-based PV cell, 25$^\circ$C)~\cite{810nmPV};
	\end{itemize}
	Moreover, we assume $\eta_{\rm pc}$, $\eta_{\rm pb}$, and $\eta_{\rm dc}$ are 100\%, because the path controller, the power buffer, and the DC-DC converter have less energy consumption~\cite{Wai2007,TPSM84624}. Therefore, the point-to-point charging efficiency $\eta$ is 20\% based on (\ref{equ:efficiency}). For the alternative scheduling algorithm, the frame time $T_{\rm f}$ is 0.2~ms. For the TDMA scheduling algorithm, the looping time (one segment) $T_{\rm g}$ is 1 sec; the total slot number of a frame, $N_{\rm s}$, is 200; and the frequency of the PWM wave is 5~kHz. Thus, the slot time $T_{\rm s}$ is 1~$\rm \mu s$ and the frame time~$T_{\rm f}$ is 0.2~ms. Each receiver is equipped with a single cell Li-ion battery whose  maximum capacity is  1000~mAh .The maximum charging power is 4.2~W, as shown in Fig.~\ref{fig:li-battery-profile}.
	
\subsection{Algorithm Comparison}

\begin{figure}
	\centering
	\includegraphics[width=3.5in]{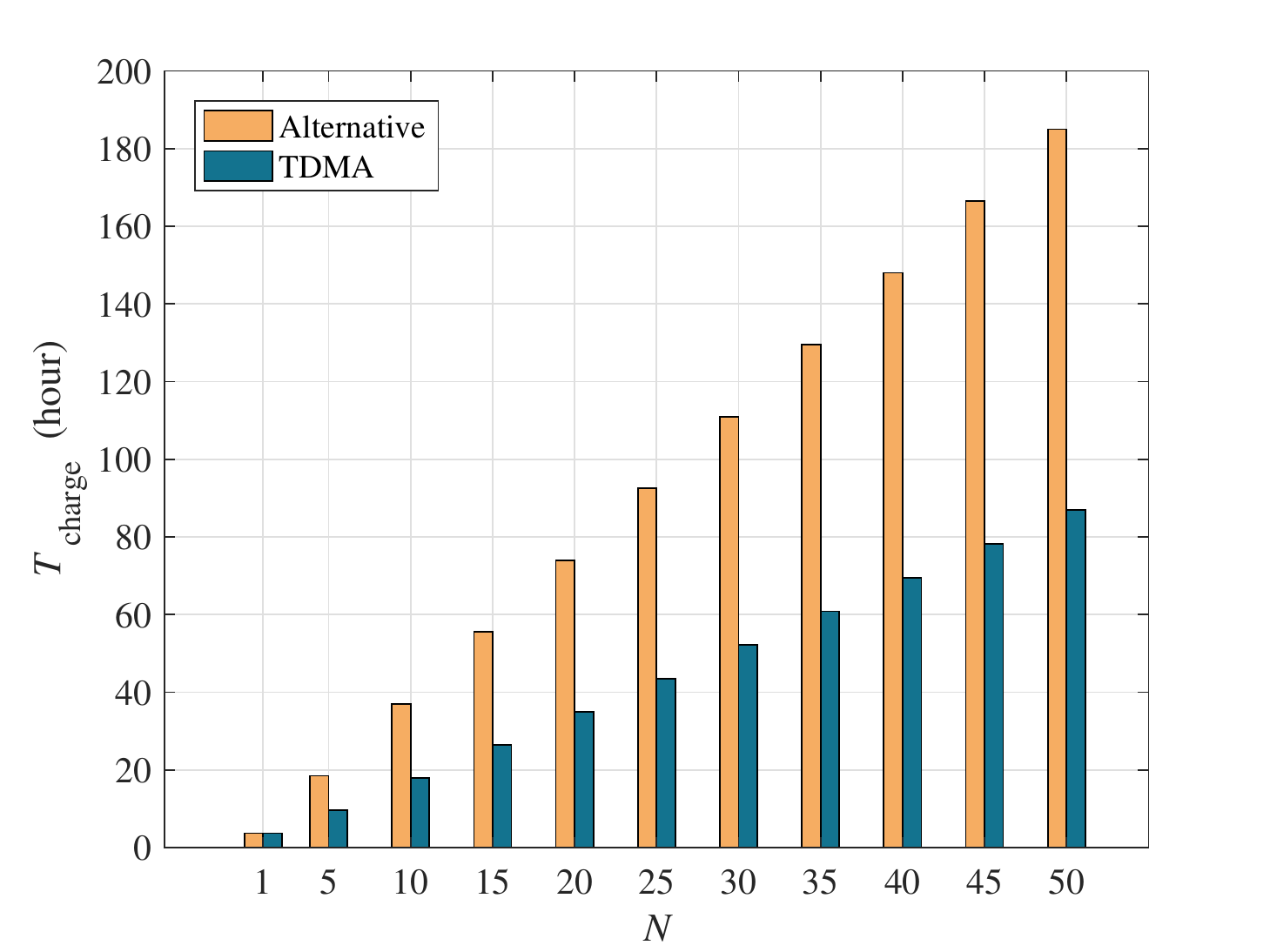}
	\caption{Charging Time  $T_{\rm charge}$ vs. Receiver Number $N$ (0 initial residual battery capacity)}
	\label{fig:0-TvsN-algorithm}
\end{figure}
\begin{figure}
	\centering
	\includegraphics[width=3.5in]{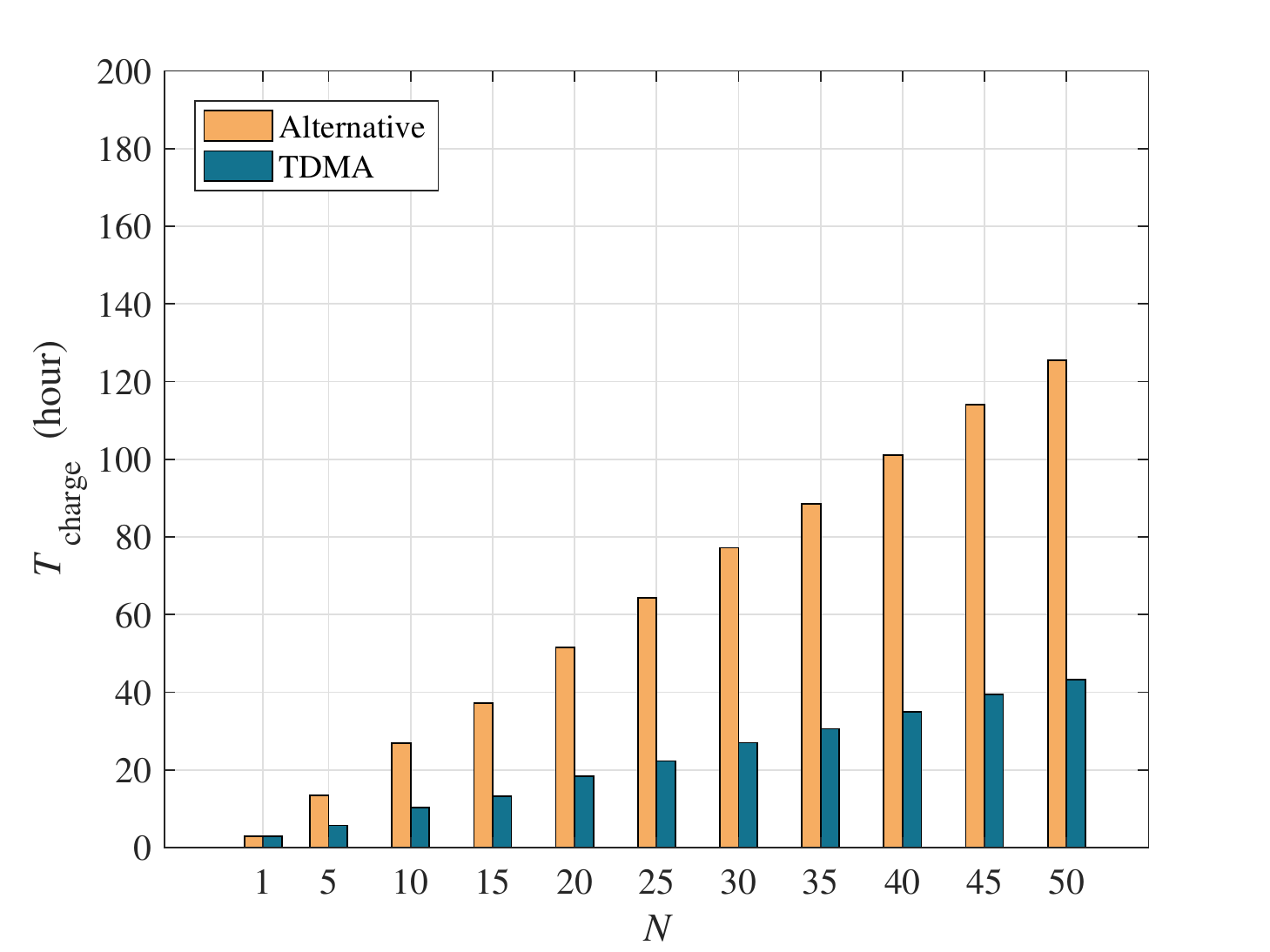}
	\caption{Charging Time  $T_{\rm charge}$ vs. Receiver Number $N$ (random initial residual battery capacity)}
	\label{fig:rand-TvsN-algorithm}
\end{figure}
\begin{figure}
	\centering
	\includegraphics[width=3.5in]{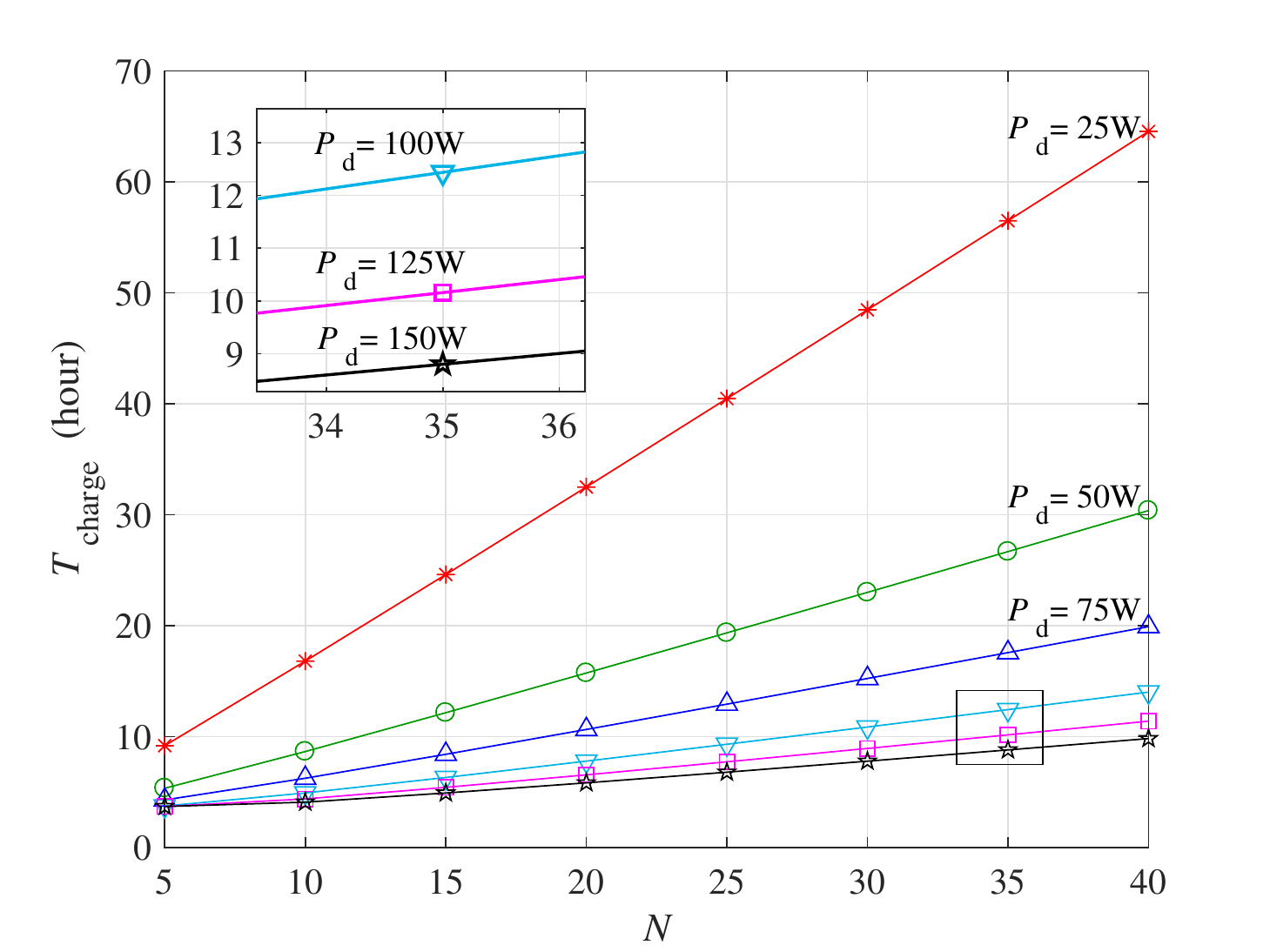}
	\caption{Charging Time $T_{\rm charge}$ vs. Receiver Number $N$ (0 initial residual battery capacity, TDMA)}
	\label{fig:0-TvsN-PP}
\end{figure}
\begin{figure}
	\centering
	\includegraphics[width=3.5in]{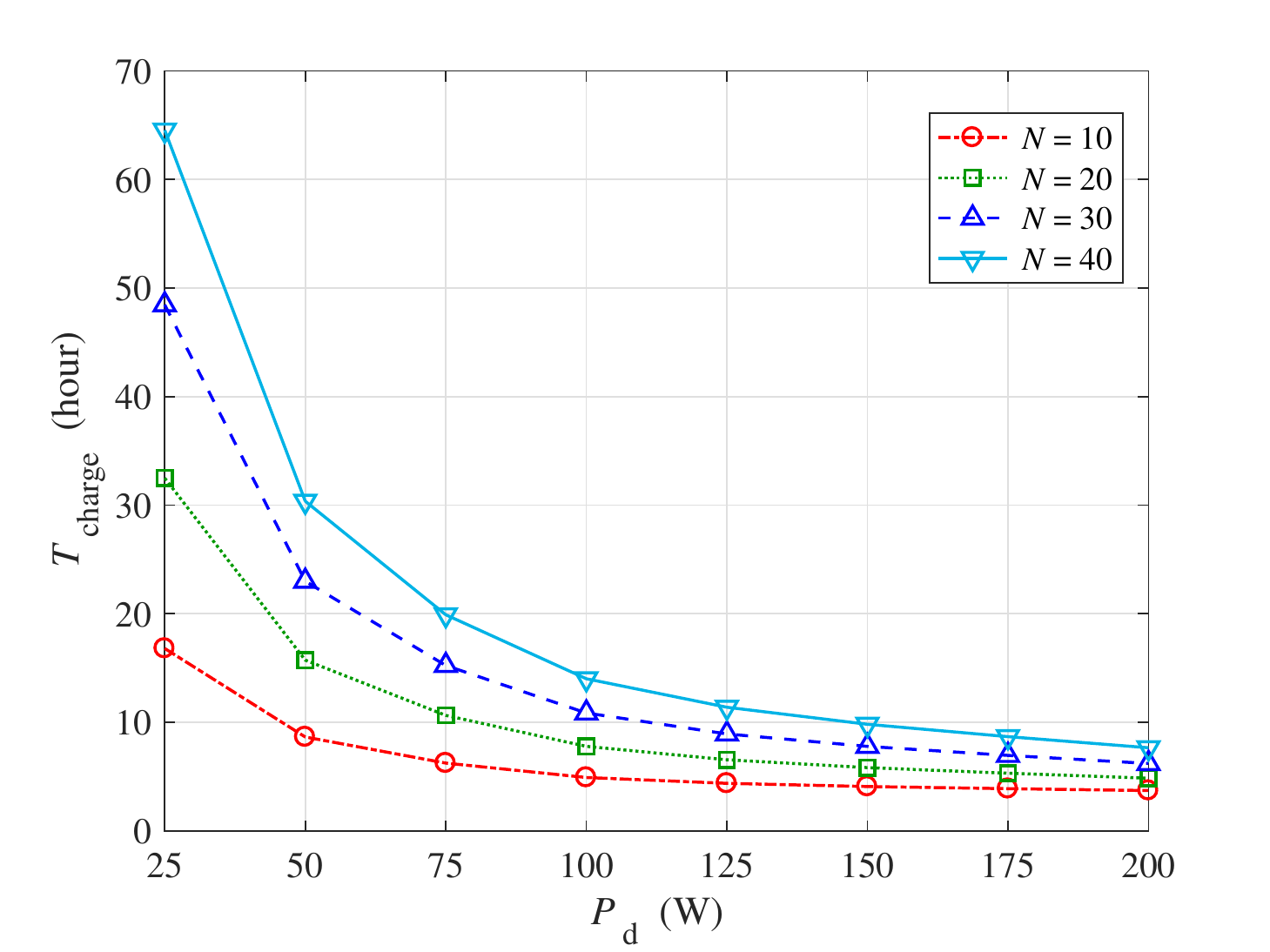}
	\caption{Charging Time $T_{\rm charge}$ vs. Driving Power $P_{\rm d}$ (0 initial residual battery capacity, TDMA)}
	\label{fig:0-TvsPP-N}
\end{figure}
 	In order to compare the efficiency between the TDMA scheduling and the alternative scheduling algorithms. The initial residual capacity of each receiver's battery is set as zero. In the alternative scheduling algorithm, the driving power is dynamic, and its maximum value is $21~{\rm W}$ calculated by $4.2~{\rm W}/20\%$, where 4.2~W is the maximum charging power according to~Fig.\ref{fig:li-battery-profile}. For fair comparison, we set the driving power $P_{\rm d}$ as 21~W for the TDMA scheduling algorithm. To compare the two algorithms quantitatively, we define 
 	\begin{itemize}
 	\item \emph{Charging time}: $T_{\rm charge}$, the time duration of charging the battery from the initial residual capacity to full capacity for all accessed receivers.
 	\end{itemize}
 
	Fig.~\ref{fig:0-TvsN-algorithm} shows the variation of $T_{\rm charge}$ with the different receiver number $N$. The relationship between~$T_{\rm charge}$ and~$N$ is close to linearity. $T_{\rm charge}$ of the TDMA scheduling algorithm is about half (46.9\% when $N=50$) of that of the alternative scheduling algorithm, except for the special situation where $N$~=~1.

	In practice, the initial residual capacity should be random rather than zero. We can assume the battery's initial residual capacity is uniformly distributed. In order to minimized the impact of randomness, the simulation runs for~10~times  to obtain the average result. Actually, we have run the simulation for more times and found less distinction on results. As shown in Fig.~\ref{fig:rand-TvsN-algorithm}, the $T_{\rm charge}$ of the TDMA scheduling algorithm is almost one third~(34.5\% when $N=50$) of that of the alternative scheduling algorithm.

	In summary, from Fig.~\ref{fig:0-TvsN-algorithm} and Fig.~\ref{fig:rand-TvsN-algorithm}, we find that the efficiency of the TDMA scheduling algorithm is greater than that of the alternative scheduling algorithm.
		
\subsection{TDMA Charging Algorithm Features}

	For the alternative scheduling algorithm, the transmitting beam power is dynamic depending on each receiver's requirement. However, for the TDMA scheduling algorithm, the transmitting beam power is fixed. Thus, the driving power $P_{\rm d}$ is customizable in the TDMA scheduling algorithm, and it can be an desired value. We evaluate here the impacts of the driving power $P_{\rm d}$, thus the transmitting beam power with the TDMA scheduling algorithm.

\begin{figure}
	\centering
	\includegraphics[width=3.5in]{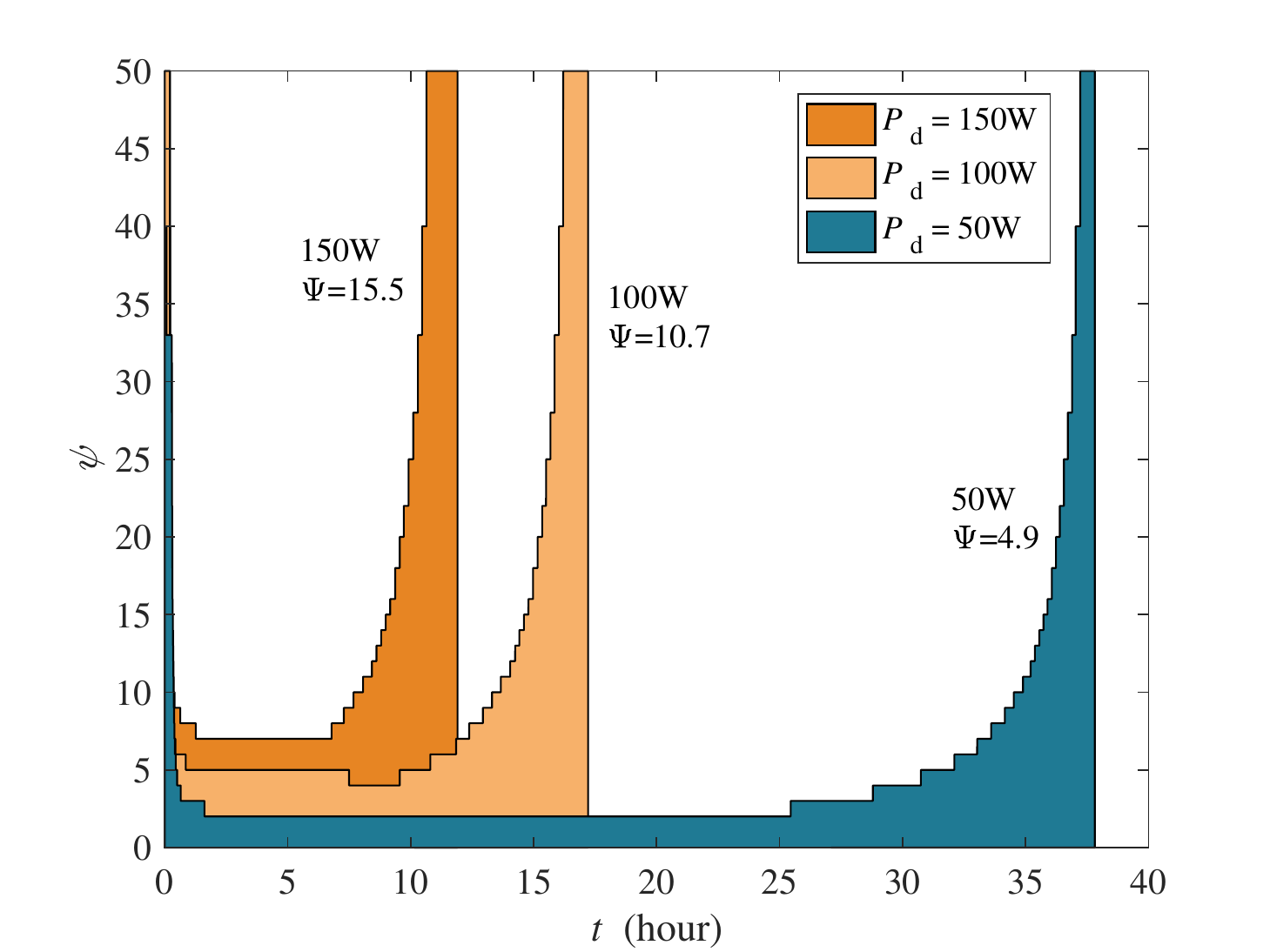}
	\caption{Multiplexing Number $\psi$ vs. Time $t$ (0 initial residual battery capacity, $\Psi$ is average multiplexing number, TDMA)}
	\label{fig:0-SNovert-PP}
\end{figure}

	We assume the initial residual capacity is zero. The receiver number $N$ varies from 5 to 40. The driving power $P_{\rm d}$ is from 25~W to 150~W. In Fig.~\ref{fig:0-TvsN-PP}, with the same $P_{\rm d}$, $T_{\rm charge}$ increases almost linearly with $N$ increasing. Moreover, we find out that the slopes of linear lines decline when $P_{\rm d}$ increasing. For example, the slope of the line  for~$P_{\rm d}$~=~50~W is about half of that for~$P_{\rm d}$~=~25~W.

	Fig.~\ref{fig:0-TvsPP-N} illustrates that $T_{\rm charge}$ decreases when $P_{\rm d}$ increasing for the different receiver number $N$. However, the impact of the small $P_{\rm d}$ on $T_{\rm charge}$ is more than that of the large $P_{\rm d}$. Furthermore, for the same $P_{\rm d}$, $T_{\rm charge}$ increases when $N$ increasing.

    Fig.~\ref{fig:0-SNovert-PP} demonstrates the variation of the multiplexing number $\psi$ during the charging period for~$N$~=~50 receivers. We consider three cases for $P_{\rm d}$~=~50~W, 100~W, and 150~W, respectively.  We find that the variation of $\psi$ can be divided into three stages: 1) initially, the $\psi$  is very large; 2) then, it goes down to a very low value and lasts for a long time; 3) at last, it rises up to a high value again. 
	
	The variation pattern of the multiplexing number~$\psi$  can be explained by two reasons. The first reason is related to the Li-ion battery charging profile (Fig.~\ref{fig:li-battery-profile}), i.e., the desired charging power $P_{\rm c}$ changes from a low level to a high level and then back to the low level during the charging period. Accordingly, the TDMA frame can contain more receivers if the desired charging powers are at the low level. Another reason is that the receivers with 0 initial residual capacity must pass the \emph{low~multiplexing~number~stage}, where $P_{\rm c}$ is high (see Fig.~\ref{fig:li-battery-profile}) and one receiver may occupy most slots in the frame. 
	
\begin{figure}
	\centering
	\includegraphics[width=3.5in]{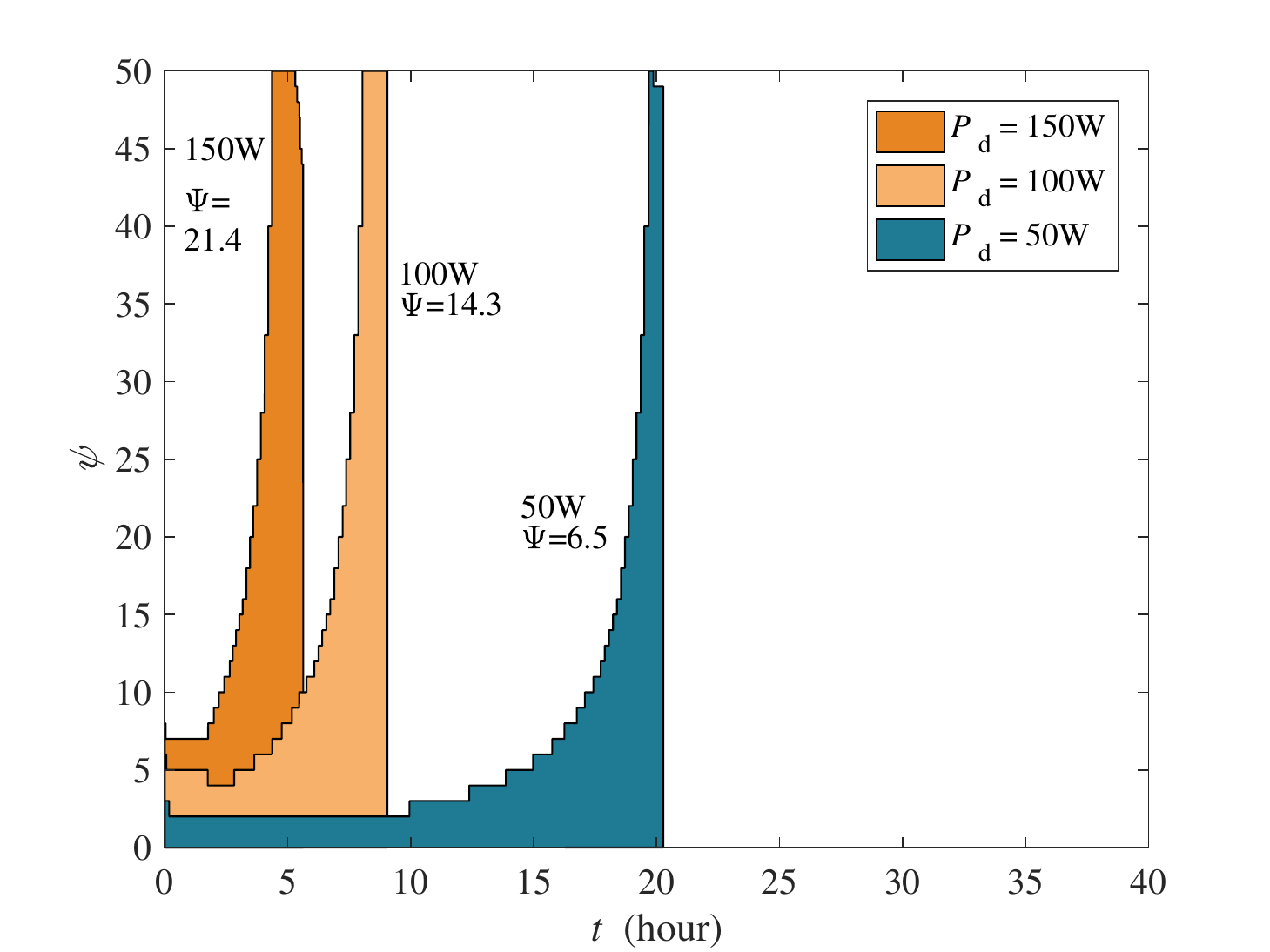}
	\caption{Multiplexing Number $\psi$ vs. Time $t$ (random initial residual battery capacity, $\Psi$ is average multiplexing number, TDMA)}
	\label{fig:rand-SNovert-PP}
\end{figure}

	Similar to Fig.~\ref{fig:0-SNovert-PP}, Fig.~\ref{fig:rand-SNovert-PP} illustrates the simulation results for the uniformly distributed initial residual capacity. For the same driving power~$P_{\rm d}$, the  average multiplexing numbers $\Psi$ in Fig.~\ref{fig:rand-SNovert-PP} are higher than that in Fig.~\ref{fig:0-SNovert-PP}. 
	
	With random initial residual capacities, it is highly probable that a receiver can be selected for the unallocated slots in the scheduling frame. Therefore, the scheduling frame can multiplex more receivers than the case with zero initial residual capacity. On the other hand, in Fig.~\ref{fig:rand-SNovert-PP}, due to the random initial residual capacity, some receivers' battery states may be at the tail of its charging profile as in Fig.~\ref{fig:li-battery-profile}. Thus, the required $P_{\rm c}$ of these receivers should be low, which may lead to the small $N_{\rm c}$. Therefore, the low multiplexing number stage in the middle of the charging period in Fig.~\ref{fig:rand-SNovert-PP} is shorter than that in Fig.~\ref{fig:0-SNovert-PP}. Moreover, the average multiplexing number $\Psi$ in Fig.~\ref{fig:rand-SNovert-PP} is larger than that in Fig.~\ref{fig:0-SNovert-PP}.

	\subsection{Summary}
	
	The TDMA scheduling algorithm has higher efficiency than the alternative scheduling algorithm. For the TDMA scheduling algorithm, the relationship between the charging time~$T_{\rm charge}$ and the receiver number~$N$ is close to linearity. The relationship between $T_{\rm charge}$ and the driving power~$P_{\rm d}$ as in Fig. 13 provides a guideline for the transmitter design. Moreover, the variation of the multiplexing number~$\psi$ reflects the efficiency. Hence, increasing the value of $\psi$ is an important approach to design more efficient TDMA charging schedulers.

\section{Conclusions}
\label{sec:conclusion}
	We propose the time-division multiple access (TDMA) charging design for multi-user wireless power transfer~(WPT) in the adaptive resonant beam charging (ARBC) system. At first, we present the point-to-point ARBC system; describe multi-user charging problem; and demonstrate the alternative charging method. Then, we introduce the principle of the TDMA charging in the ARBC system and depict its digitalized implementation. Next, we propose the TDMA-based scheduling algorithm. Finally, we compare the efficiency of the alternative and TDMA scheduling algorithms by simulation. In summary, the TDMA charging enables multi-user WPT with continuous battery charging current. Furthermore, it provides the transmitter with flexible output capability, because the maximum output power has no limitation. The simulation results show that the TDMA scheduling algorithm has high efficiency, as the total charging time is almost half (46.9\% when charging 50 receivers) of that of the alternative scheduling algorithm.
	
	In the future, we will investigate the efficiencies of the circuit components and the dynamic property of Li-ion battery. Moreover, flow control and the scheduler with quality of service (QoS) can be further studied.
	
\section*{Acknowledgements}
	We would like to thank my friends and colleagues in our laboratory. We would like to thank Hao Deng and Wen Fang for their valuable suggestions. We would also like to thank Aozhou Wu for his polishing of the figures in this paper.


%

\appendices
%



\ifCLASSOPTIONcaptionsoff
  \newpage
\fi



%
\bibliography{TDMA-Charging}
%
%

%

%
%







\end{document}